# E-QED: Electrical Bug Localization During Post-Silicon Validation Enabled by Quick Error Detection and Formal Methods

Eshan Singh, Clark Barrett, Subhasish Mitra

Stanford University, USA
{esingh, clarkbarrett, subh}@stanford.edu

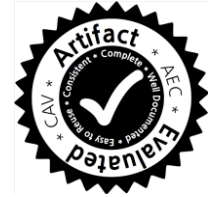

**Abstract.** During post-silicon validation, manufactured integrated circuits are extensively tested in actual system environments to detect design bugs. Bug localization involves identification of a bug trace (a sequence of inputs that activates and detects the bug) and a hardware design block where the bug is located. Existing bug localization practices during post-silicon validation are mostly manual and *ad hoc*, and, hence, extremely expensive and time consuming. This is particularly true for subtle electrical bugs caused by unexpected interactions between a design and its electrical state. We present E-QED, a new approach that automatically localizes electrical bugs during post-silicon validation. Our results on the OpenSPARC T2, an open-source 500-million-transistor multicore chip design, demonstrate the effectiveness and practicality of E-QED: starting with a failed post-silicon test, in a few hours (9 hours on average) we can automatically narrow the location of the bug to (the fan-in logic cone of) a handful of candidate flip-flops (18 flip-flops on average for a design with ~ 1 Million flip-flops) and also obtain the corresponding bug trace. The area impact of E-QED is ~2.5%. In contrast, determining this same information might take weeks (or even months) of mostly manual work using traditional approaches .

## 1    Introduction

For complex integrated circuits (*ICs*), difficult design flaws (bugs) increasingly escape pre-silicon design verification and are only detected during post-silicon validation when manufactured ICs are extensively tested in actual system environments [Foster 15]. Design bugs can be broadly classified into *logic bugs* that are caused by (logic) design errors and *electrical bugs* that are caused by unexpected interactions between a design and its electrical state. Examples include errors introduced by crosstalk, power-supply noise, thermal effects or process variations. Traditional pre-silicon verification is slow; but, more importantly, it is generally incapable of detecting electrical bugs that appear only after ICs are manufactured. **This paper focuses on electrical bugs.**

Typical post-silicon validation involves: 1. bug detection by applying a variety of test stimuli (e.g., random instruction tests, end-user applications) at various voltage, temperature, and clock frequency corners; 2. *bug localization* which identifies a *bug trace* (a sequence of inputs that activates and detects the bug) and a hardware design block where the bug is located; and, 3. bug fixing using techniques such as software patches, clock frequency/operating voltage selection, or silicon respin. Existing post-

---

This work is supported in part by DARPA and the Semiconductor Research Corporation (SRC).

silicon validation and debug practices are mostly manual and *ad hoc*, and, hence, very expensive [Mishra 17]. The effort to localize bugs from observed system failures (e.g., crashes, output errors) dominates the overall cost [Dusanapudi 15, Friedler 14, Nahir 14]. For example, it might take weeks (or even months) of (manual) work to localize a single bug [Dusanapudi 15, Reick 12, Vermeulen 14].

Post-silicon bug localization is difficult because of long error detection latencies. *Error detection latency* is the time elapsed between when a test activates a bug and creates an error and when that error manifests as an observable failure (e.g., system crash). Error detection latencies of difficult bugs can exceed several millions or even billions of clock cycles [Hong 10, Lin 14]. It is extremely difficult to trace that far back into the history of system operation for complex ICs. In addition, IC design size and complexity pose major challenges. Full-chip simulation to obtain expected responses (for various internal states, not just software-visible states) is several orders of magnitude slower than actual silicon and may be impractical. Formal analysis and Boolean Satisfiability techniques can be severely limited by design size. System-level failure reproduction, which involves returning the system to an error-free state and re-running the system (perhaps with some modifications) to reproduce the "exact" failure, is difficult (due to non-deterministic behaviors such as asynchronous I/Os and multiple-clock domains). In order to limit the number of cycles that must be traced and analyzed during bug localization, techniques such as Quick Error Detection (*QED* [Lin 14]) that ensure short error detection latencies are crucial. An overview of existing bug localization approaches is presented in Sec. 4.

New techniques are essential to overcome post-silicon bug localization challenges. There have been some recent publications that address detection and localization of logic bugs (e.g., [Lin 15]) during pre-silicon and post-silicon validation. Here, we present E-QED, a new technique to automatically localize electrical bugs during post-silicon validation and debug. Key features of E-QED are: 1. It is broadly applicable to most digital designs. 2. It can localize electrical bugs inside processor cores as well as in uncore components (interconnection networks, cache controllers, memory controllers) that occupy large portions of System-on-Chip (*SoC*) designs. 3. It doesn't require manual intervention during design or during post-silicon validation and debug. 4. It doesn't rely on full system-level simulation. 5. It scales to large designs.

We demonstrate the effectiveness and practicality of E-QED using OpenSPARC T2, a 500-million-transistor open-source SoC. Our results (details in Sec. 3) show that: 1. E-QED correctly and automatically localizes electrical bugs in a few hours (between 7-13 hours). Such bugs would generally take weeks (or even months) of manual work to localize using traditional approaches. 2. E-QED achieves very fine-grained electrical bug localization. For each localized electrical bug, using formal analysis, E-QED automatically generates a small list of candidate flip-flops (*FFs*) that might have captured error(s) caused by the bug. For the OpenSPARC T2 SoC with ~1 Million FFs, E-QED automatically localizes electrical bugs to only 18 candidate FFs on average. Thus, E-QED achieves a *localization factor* (total number of FFs in a design divided by the total number of candidate FFs that an electrical bug is localized to) of over 50,000. 3. For each localized electrical bug, E-QED automatically generates a short bug trace using formal analysis. 4. E-QED incurs only a small area overhead (~2.5% for OpenSPARC T2) and has practically no clock-speed impact. 5. E-QED enables flexible trade-offs between area overhead, electrical bug localization granularity, and bug localization runtime. For example, with 1.5% area overhead (vs. 2.5%) the average FF candidate count increases only by a factor of 3 (i.e., E-QED still achieves a localization factor of four orders of magnitude).

E-QED uses the following three steps that work together in a coordinated fashion to overcome electrical bug localization challenges: 1. Low-cost hardware structures called E-QED signature blocks are automatically inserted during the design phase (Sec. 2.1). 2. QED tests that achieve short error detection latencies are run during post-silicon validation.[1] 3. Formal techniques are used to reason about the signatures collected by the E-QED signature blocks, automatically localizing bugs to a handful of candidate FFs (Sec. 2.2-2.4).

*Motivating Example*

Consider the following electrical bug example for the OpenSPARC T2 SoC [OpenSPARC], shown in Fig. 1. Suppose that a bug occurs during the following sequence of events: 1) Processor core 0 writes the value 0 to the crossbar (in order to store the value 0 in address [C]); 2) the entry corresponding to address [C] in L2 cache bank 0 is updated with value 0 (from the crossbar); 3) in response to a request to load from address [C] by processor core 3, the cached value corresponding to address [C] is written into the crossbar; 4) when this value passes through the crossbar arbitration logic, 5) an electrical bug causes a single-bit error to be captured in the output register of the crossbar arbiter; 6) the corrupted value is loaded into core 3; and then, 7) the corrupted value is detected by a QED test.

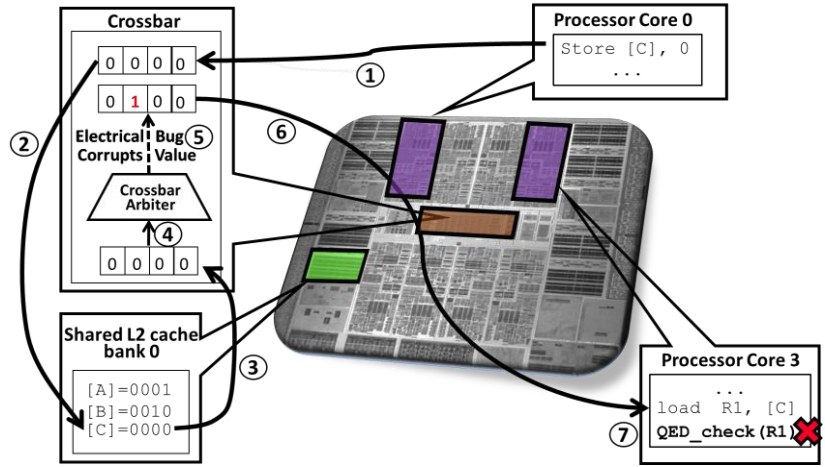

**Fig. 1**: An example of an electrical bug corrupting the value corresponding to address [C] stored by processor core 0 to L2 cache bank 0 (steps 1-2) as it passes through the arbitration logic of the crossbar (steps 3-5) while being loaded by processor core 3 (step 6). The bug is detected by a QED test (step 7).

---

[1] We use Quick Error Detection (*QED*) tests that typically achieve error detection latencies of 1,000 clock cycles or fewer [Lin 14]; however, our approach can work with other tests that achieve similar error detection latencies.

Note that without QED, the error may not be detected by the post-silicon validation test or the error detection latency can be millions of clock cycles; e.g., the error may be detected during an end-result check (which checks for expected output values upon test completion). Using a QED test, the error detection latency improves (i.e., reduces) significantly to only a few hundred cycles. This drastically reduces the amount of data that needs to be analyzed to localize the bug. However, upon error detection by the QED test in processor core 3, it is impossible to <u>directly</u> determine where the error actually occurred and consequently how to localize the bug to a specific design block. An error in the datapath during any of steps 1-6 (in processor core 0, the crossbar, L2 cache bank 0, or processor core 3) would have been detected in the same way by the QED test running on processor core 3 (using the QED check in Fig. 1, step 7).

E-QED automatically localizes this bug not only to the crossbar (containing 40,000 of the nearly 1 million FFs in the design), but to a subset of 8 candidate FFs within the crossbar that could have captured this single-bit error. The 8 candidates include the FF in the output register that did capture the actual error. E-QED also provides a 544-Kbit bug trace, the sequence of inputs to the crossbar that triggered the bug during post-silicon validation. The total runtime required to obtain this result by running E-QED is 488 minutes, and the hardware area overhead is 2.5%.

In contrast, traditional post-silicon bug localization approaches would require significant manual effort, additional hardware (e.g., trace buffers, details in Sec. 4) with significantly higher area overhead, or both. For example, even if the bug was detected quickly using a QED test, saving a full trace of all inputs and outputs of the crossbar alone for just 1,000 cycles would require over 34 Mbits of data.

The rest of the paper is organized as follows. Section 2 presents our new E-QED technique. Results are then presented in Sec. 3, followed by related work in Sec. 4. We conclude in Sec. 5.

## 2  Electrical Bug Localization using E-QED

E-QED relies on several steps that are summarized in Fig. 2. During the chip design phase, E-QED signature blocks are automatically inserted (details in Sec. 2.1). These E-QED signature blocks are used during post-silicon validation to capture and compress the logic values of selected signals (*signatures*). During post-silicon validation, a suite of tests is run—it is crucial to run tests with short error-detection latencies (e.g., QED tests [Lin 14]). When an error is detected by such a test, the test is immediately halted, and all the captured signatures are scanned out (using on-chip scan chains [Abramovici 90]). In the last phase, formal analysis, the collected signatures are analyzed by a Bounded Model Checking (*BMC*) tool to first identify which design block produced the error(s), and then to find the FFs in that block that could have captured the error(s), and finally to narrow this list even further by checking for consistency with signatures captured by neighboring design blocks. We explain each of these steps in more detail below.

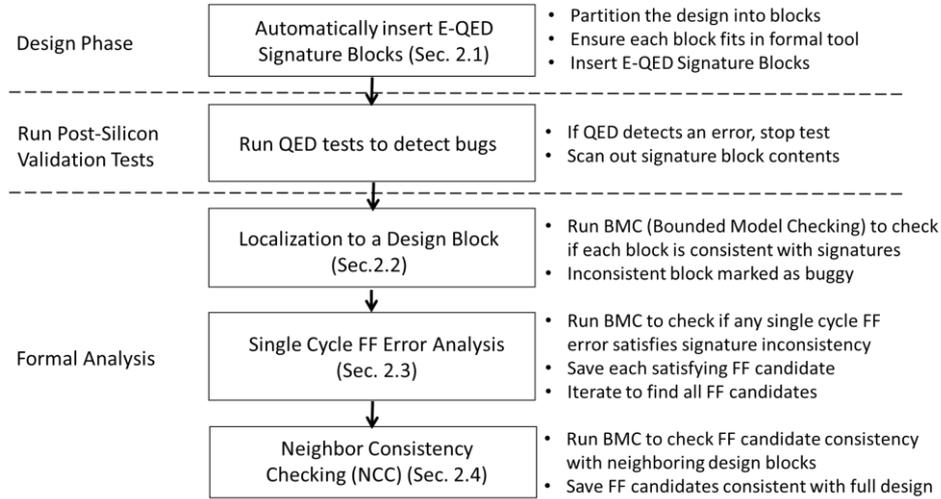

**Fig. 2**: An overview of E-QED.

### 2.1 Automatic Insertion of E-QED Signature Blocks

E-QED relies on being able to use a formal tool to perform bug localization after error detection by a post-silicon validation test. If the entire design can be loaded into the formal tool and analyzed, then this requirement can be satisfied by externally tracing the design's inputs and outputs during post-silicon testing (i.e., by saving all the signals at the external design interface). In this case, no additional internal hardware is required. However, most designs (particularly large SoCs) cannot be analyzed by existing formal tools without being partitioned into smaller blocks. In order to be able to analyze these smaller design blocks, we insert additional hardware (*E-QED Signature Blocks*) to capture logic values of signals at the boundaries of these smaller design blocks.

**Design Partitioning.** As mentioned above, unless it is possible to fit the full design into a formal tool, the design must be divided into smaller blocks. We use a simple algorithm that builds a list of design blocks by recursively descending through the design hierarchy. At each step, the current design block is tested to see if it can be loaded into the formal tool. If so, the recursion terminates and the block is added to the list. The result is a partition of the design into design blocks, each of which can be analyzed by the formal tool. Next, input and output signals for each block are grouped into *interfaces* as follows. For each design block A, the set of output signals driven by A and captured by the same design block B (or the same set of design blocks in the case that the signals fan out to more than one design block) are grouped into a single interface. In addition, if a design block C receives inputs from a design block in a different clock domain, then all such input signals for design block C are also grouped into a separate interface. Each interface gets a single E-QED Signature Block (explained next) as illustrated in Fig. 3. As explained above, the primary inputs and outputs are traced externally.

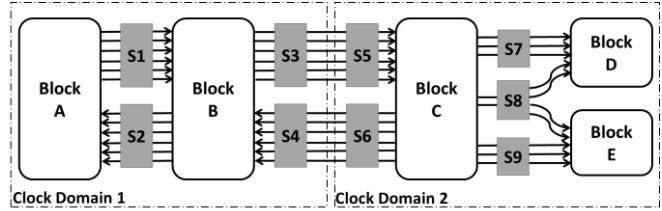

**Fig. 3**. Insertion of E-QED Signature Blocks. A and B use the same clock, so they can share signature blocks S1 and S2 at their interfaces. Since C is in a separate clock domain, the interfaces between B and C require separate signature blocks within each clock domain. Each signal from C captured in signature block S8 fans out to both D and E.

In our evaluation in Section 3, we explored one further enhancement beyond this algorithm. We observed that uninitialized memory arrays contributed too many degrees of freedom during formal analysis. To overcome this issue, we inserted additional signature blocks. Specifically, we added signature blocks to the signals between each cache bank and the cache controller logic and to the signals between the instruction cache and the instruction fetch unit for each thread on each processor core. This provided additional information on the data values being stored and loaded in the caches and the instructions being executed by the processors. As shown in Fig. 12 (in Sec. 3.2), this significantly improves the precision of E-QED.

**E-QED Signature Block Design.** E-QED Signature Blocks store logic values called signatures that represent a (lossy) compression of the sequence of logic values for a set of signals over time. We use multiple-input signature registers or *MISRs* (which have been extensively used for circuit test response compaction [Saxena 97]) for this purpose. A MISR is a shift-register in which certain bits are XORed and fed back into the first bit, and, at the same time, a set of input signals are XORed with the values being shifted. An example of a 6-bit MISR is shown in Fig. 4.

If a MISR has been operating for $N$ cycles since reset, we refer to the number $N$ of captured cycles as the *capture window* of the MISR. For post-silicon bug localization, it is crucial that the capture window is long enough to include the point when a bug gets activated and an error is created. However, the capture window must also not be too long, since the design behavior over the entire window is analyzed using the formal tool (see Sec. 2.2-2.4). If $N$ is too large, the formal tool will fail as the unrolled design grows too large to analyze. This is why tests with short error detection latencies are necessary, and QED in particular enables our E-QED approach. For most cases, QED ensures that capturing the last 1,000 cycles prior to error detection is sufficient to also capture the point of bug activation. In our experiments (Sec. 3), we set $N$ to 1,024. Since post-silicon validation tests run longer than 1,024 cycles, each MISR will be reset to a known state periodically to maintain the capture window length (1,024 cycles in this case).

In choosing the MISR size, we select the number of bits $K$ to be at least equal to the number of signal bits being captured ($M$). $K$ should also be large enough such that the MISR has more states than the capture window length (i.e., $K$ must be greater than $\log_2(N)$). For $M$ signal bits over $N$ cycles, a complete trace would require $M*N$ bits. Thus, if $K$ is $M*b$ (can be viewed as $b$ MISR bits per input signal), the parameter $b$ decides the compression ratio $N/b$. Based on empirical analysis (discussed in Sec. 3.2), we used $b = 8$ for all the blocks in the OpenSPARC T2 design except for inside the processor core, where we used $b = 4$. After selecting the MISR size, the choice of which bits to use as feedback can be made based on extensive existing work which optimizes for various good properties of MISRs (e.g., [Bardell 87]).

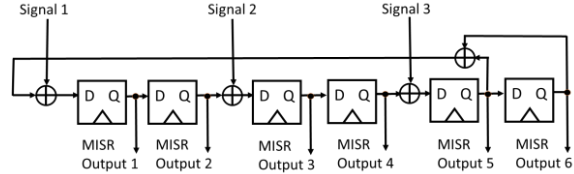

**Fig. 4**: An example of a 6-bit Multiple Input Signature Register (MISR) with 3 input signals and feedback generated from bits 5 and 6.

As mentioned above, we periodically reset each MISR to a known state (to mark the start of a capture window). However, this raises the possibility that the error might be captured right after the MISR is reset, in which case the number of cycles captured will be much less than our target $N$ (and the point when the bug got activated may not be included in the capture window). To avoid this situation, in our *E-QED Signature Block* (shown in Fig. 5) design, we use two MISRs (operating in parallel) for each signature block. The two MISRs are paired with a counter, and both are reset (the MISRs to a defined starting state and the counter to 0) when the design is powered on with a global reset. After that, the counter repeatedly counts to $2N$, alternately resetting one of the MISRs (to its starting state) every $N$ cycles (waveforms in Fig. 6). This ensures that at the instant an error is detected, at least one of the MISRs has captured a signature covering at least $N$ (and no more than $2N$) cycles. By setting $N$ to a value greater than the expected error detection latency (e.g., 1,000 cycles for QED tests), we can ensure that with high likelihood, at least one signature covers the entire period between bug activation and error detection. The counter is sized to $C$, the ceiling of $\log_2(2N)$. The logic values in the MISRs and the counter can be scanned out when an error is detected.

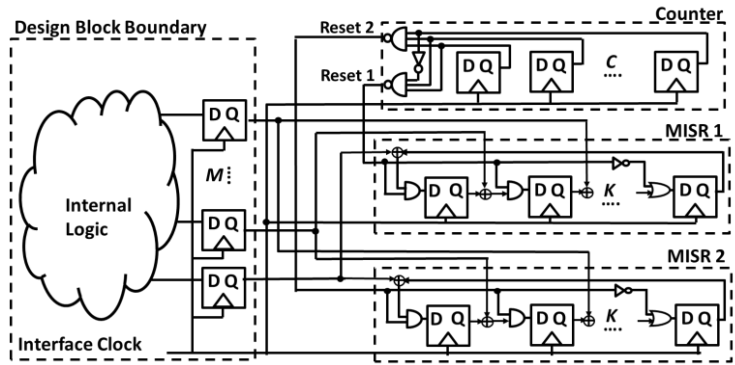

**Fig. 5**: The complete E-QED Signature Block design, with $M$ interface signals generating a $K$-bit signature and the reset logic shown with the FFs of a $C$-bit counter (detailed counter implementation and MISR feedback not shown).

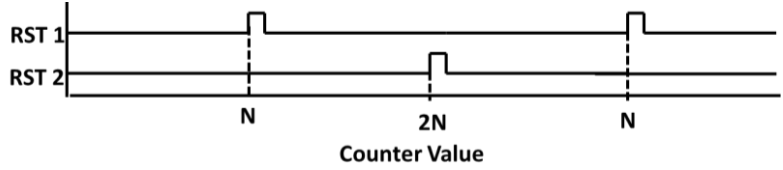

**Fig. 6**: Reset of MISR 1 and MISR2, each resetting every *2N* cycles. Note that at any instant, one MISR has captured data covering at least *N* cycles.

## 2.2 Bug Localization to a Design Block

We now explain how our design partitioning approach in Sec. 2.1 supports the first step in the electrical bug localization process: isolating an error detected by a post-silicon validation test to a specific design block. Following the detection of an error by QED tests, the logic values captured in the counters and MISRs are scanned out from each E-QED signature block. A formal analysis of the signature values is then performed using BMC [Clarke 01]. BMC works by taking a model of the system (e.g., Verilog RTL), unrolling it some fixed number of time steps, and employing iterative time frame expansion to model sequential operation over this window of time. Then, automated tools such as Boolean Satisfiability solvers are used to check whether a set of constraints on the unrolled design is consistent. Constraints can be placed on any part of the unrolled design, commonly including the initial state, the inputs, and the final state.

Recall that when an error is detected, at least one of the two MISRs in each E-QED signature block has captured at least *N* cycles. Let *T* be the larger of the two capture windows (obtained from the counter in the E-QED signature block). For each design block *B* to be analyzed, we set up a BMC problem over *T* unrolled cycles (time frames), where *T* is the larger of the two windows captured by the E-QED Signature Blocks for the interfaces associated with *B* (i.e., signals that are either inputs or outputs for *B*).[2] In the BMC problem, we constrain the appropriate MISR (the one capturing *T* cycles) in each signature block to be equal to its scanned-out value in the final state following error detection, and also to be equal to its reset value at the beginning of the capture window (*T* cycles before the final state). This corresponds to whether the design can satisfy the following property:

```
MISR reset = 1
##<T>
Input MISRs = <Input Signatures> && Output MISRs = <Output Signatures>
```

If the BMC tool is able to solve these constraints (note that, the initial state of the design block is not specified and so can be chosen arbitrarily by the BMC tool), we conclude that no bug occurred in this particular design block during the captured cycles. The solution produced by BMC also provides one possible error free *T*-cycle trace for the design block consistent with the signatures. On the other hand, if BMC is unable to find a solution, this means that the actual signals observed by the E-QED signature blocks are not consistent with the design logic, indicating that an electrical bug must have occurred within the block.[3] Because this process identifies the buggy design block, it can already produce a significant degree of bug localization. Note that, it is theoretically possible for all design blocks to pass the BMC test, either because (i) the bug triggered (and produced error) outside the capture window; or (ii) some valid trace shares the same set of signatures as the actual captured buggy trace (because of lossy compression by E-QED signature blocks); or (iii) the symbolic initial states (corresponding to internal flip-flops in the design block where E-QED signature blocks are not inserted) introduce too many degrees of freedom. However, in our simulations based on injected errors (see Sec. 3), using QED tests and $N = 1024$, we did not encounter any such scenario: we always found exactly one block that failed the BMC test.

---

[2] We assume that all interfaces for a design block are in the same clock domain and thus all have the same value of *T*. This assumption can easily be satisfied by modifying the design partitioning algorithm to continue its recursive descent if the current design block uses more than one clock domain.

We illustrate the analysis above through a simple example, shown in Fig. 7. The inputs and outputs are both compressed using the 6-bit MISR shown in Fig. 4. For the Input MISR, its inputs, signals 1, 2, and 3 (corresponding to Fig. 4), are B, A, and 0 respectively. For the Output MISR, signals 1, 2, and 3 are Z, Y, and X respectively. During cycle 1, the MISRs are at their initial state (000001) and the circuit inputs A and B are 1 and 0. The initial outputs at cycle 1 are the contents of F6, F7 and F8 (011), and they are compressed into the output MISR on the next clock edge. A transient electrical bug occurring somewhere in the circuit results in the output X having the incorrect value 1 at cycle 5. At cycle 6, the error is captured in the output MISR signature, making it 110010 (instead of the correct output MISR signature 110000). A BMC run for this design over the 6 cycles (based on the captured output MISR signature 110010 and the captured input MISR signature 111010) reports unsatisfiable, indicating that an error has occurred in the circuit[4].

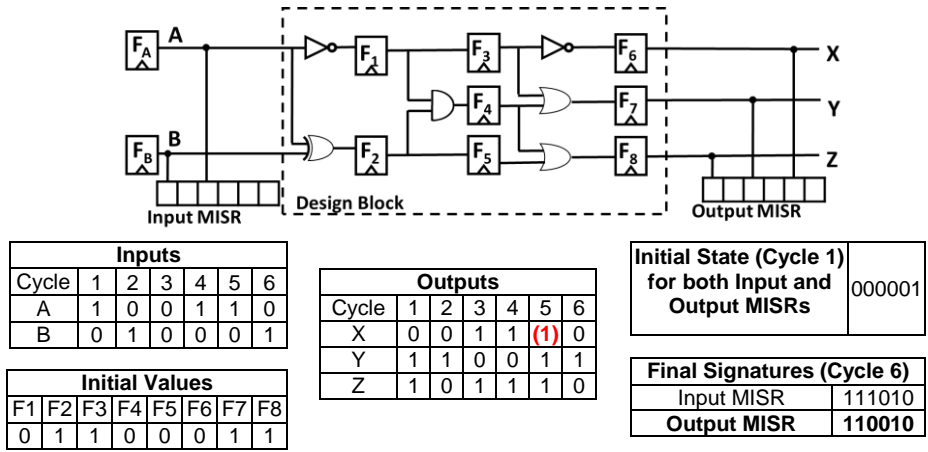

**Fig. 7**: A simple design block, a sequence of inputs and outputs, the initial values in the FFs and MISRs, the final signature captured by the Input MISR and the expected and captured (corrupted) signature in the Output MISR.

---

[3] As an optimization, in this case, we also run the BMC analysis with the smaller value of $T$ for the block. If this analysis is also inconsistent, it provides a much shorter window in which the bug occurred.

[4] E-QED uses symbolic initial values for FFs. To keep this example simple, we set the initial values of the FFs as shown in Fig. 7.

## 2.3 Single Cycle FF Error Analysis with Bounded Model Checking

After the electrical bug has been localized to a single block (recall that in Sec. 2.1, we choose block sizes based on what fits within the BMC tool), E-QED performs further localization within the block by using an error model. In this paper, we use a single-cycle FF error model, which assumes that the electrical bug causes a transient error that affects one (arbitrary) flip–flop in the design during a single (arbitrary) cycle. Such an error model has been commonly used in the literature (e.g., [McLaughlin 09]) for electrical bugs. To demonstrate how error analysis can find a list of candidates, one of which corresponds to the FF and cycle that captured the single cycle error due to an electrical bug, consider the earlier example of Fig. 7 presented again in Fig. 8.

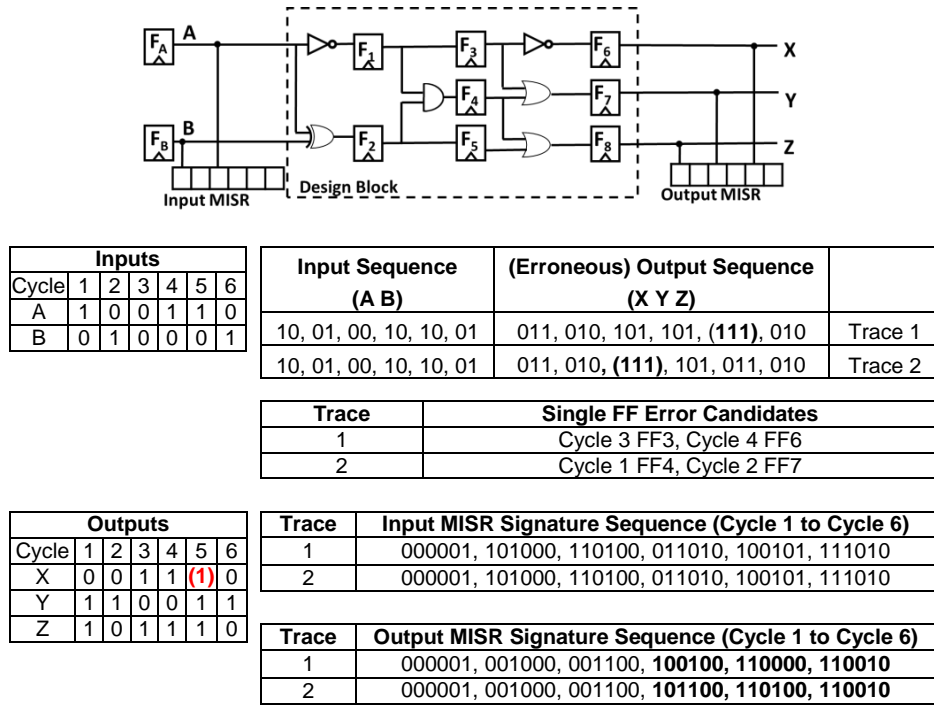

**Fig. 8:** The results of single cycle FF error analysis for the example from Fig. 7. Each candidate (cycle and FF pair) has the property that if an error occurs at the input of the candidate FF at the candidate cycle (captured in the following cycle), it would result in the captured Output MISR signature at cycle 6. Note that the candidates that generate Trace 1 match the actual sequence but the candidates that generate Trace 2 are aliased candidates in that they correspond to a different output sequence but result in the same final output MISR signature.

For simplicity, let us first assume that we can directly observe all of the inputs and outputs on every cycle (we will consider the case where we have only the MISR signatures next). Recall that the expected output at cycle 5 is XYZ = 011, but we observe: XYZ = 111; i.e., the value of X is incorrect. In this simple example, there are only three FFs along the path to the X output; $F_1$, $F_3$ and $F_6$. According to our error model, an electrical bug could have caused an error in any one of them, and all three could have potentially caused the observed error at X; so there are three possible error scenarios. In the first, the error causes $F_6$ to capture a 0 instead of a 1 in cycle 4 (due to an electrical bug along the path from $F_3$ to $F_6$). In the second, the error causes $F_3$ to capture a 0 instead of a 1 in cycle 3 (due to an electrical bug along the path from $F_1$ to $F_3$). This error then propagates to $F_6$ and then to X. In the third, the error causes $F_1$ to capture a 0 instead of a 1 in cycle 2 (due to an electrical bug along the path from $F_A$ to $F_1$). But notice that in this case, the input to $F_4$ in cycle 3 would then change to 0, and crucially, the input to $F_7$ in cycle 4 would also change to 0, and therefore output Y would be 0 in cycle 5. Since the error did not affect Y, this third scenario is not consistent with the observed outputs, leaving only errors in FFs $F_3$ and $F_6$ as viable candidates. Importantly, observe that, given our error model, this is an exhaustive list of the potential sources of the error observed at X in cycle 5.

Now, consider the case where we do not have direct access to the input or output signals, but only to the compressed 6-bit MISR signatures. Fig. 8 once again lists the captured Output MISR signatures in sequence ending at cycle 6. The two candidate FFs that were identified as possible sources of the error above both generate the captured erroneous output signature by recreating the exact output sequence (corresponding to Trace 1) that was captured due to the error. However, there are also other traces that can be created with different candidates that end with the same final signatures in the MISRs. We call these *aliased* FF candidates and traces. An example is shown as Trace 2 in Fig. 8. Aliasing occurs due to lossy compression in the MISRs: errors introduced into the output MISR at different times can result in the same final signature. Observe from the output sequences in Fig. 8 that for Trace 2, the erroneous candidate FFs cause an output error at Y in cycle 3. This error shifts through the output MISR (as shown in the MISR sequence for Trace 2) to still match the final captured output MISR signature. Since in practice we do not have the actual output signals, but only work with the MISR signatures, aliasing can increase the number of candidate FFs and traces returned by the formal analysis, reducing the resolution of the bug diagnosis. We discuss a technique for significantly improving (i.e., reducing and potentially eliminating) these aliased candidates in the next section.

We use BMC to systematically compute all candidate FFs consistent with the signature values for the interfaces of the buggy design block. To do this, we first modify the design (this modification is only to support the formal analysis at this step; the actual manufactured design is **not** affected). In the modified design, a two-input multiplexer is added before the data input of every FF in the design (Fig. 9(a)). The inputs to the multiplexer are the original data input to the FF and an inverted data input value to help inject an error as needed. By setting the select line of the multiplexer, an error is captured in the FF, matching the model for electrical bugs described earlier.

The select lines are all controlled by a decoder, allowing an error to be injected into a single FF on exactly one cycle as shown in Fig. 9(b). The decoder inputs are left unconstrained and provided to the BMC tool to control. The least significant output bit is not used, so an input combination of all 0s to the decoder generates an output of all 0s (for decoder output bits $1…R$), meaning no error is injected into the design on that cycle. An additional FF (Fig. 9b) is added to ensure that only one single FF in the design block is chosen to have an error for only one cycle during a single BMC run. This additional FF (in Fig. 9b) is initially reset to 0 (reset is asserted at the start of BMC) and is then set to 1 (and remains 1) once the decoder produces a non-zero value on output bits 1...*R*. Once the value of this FF is 1, it forces the "Error Select" signals to all be 0 for all remaining cycles. Thus, the BMC, in its attempt to satisfy the error trace, can only inject an error into a single FF in the design, and only in one cycle.

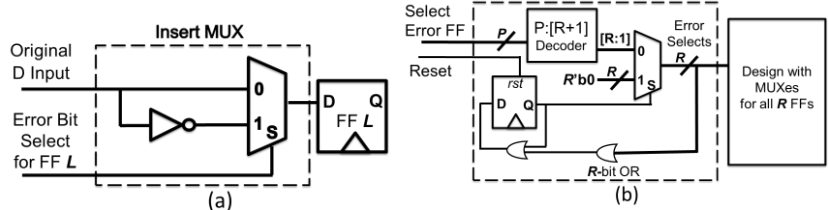

**Fig. 9:** (a) The inserted multiplexer allows an error to be injected into FF *L*. (b) control logic to inject an error during one cycle into the design when the BMC provides a FF index as input "Select Error FF" to the decoder. No error is injected when P is all 0s. The first time a non-zero input is provided, exactly one of the bits of the R "Error Select Lines" is set, causing the corresponding FF to capture an error during the next cycle.

As before, a BMC problem is set up with an unroll limit of *T* cycles, and constraints are added corresponding to the reset and final signature values. This time, however, the BMC tool can additionally control the decoder so that it can attempt to satisfy the input and output constraints by injecting a single error into a FF in the design block sometime during the *T* cycles. In this way, it finds a candidate FF which could (potentially) be the source of the erroneous bit flip (bug) that caused the error. Of course, it is possible that there is more than one way to achieve this. Therefore, the BMC tool is run repeatedly (each time with an added constraint to rule out all previously found FF candidates) until no more FFs can be found.

### 2.4  Neighbor Consistency Checking

As explained in the example in Sec. 2.3 above, a consequence of using compressed signatures to constrain BMC is that aliasing can occur: different candidates for a single FF error can lead to the same output signature. Aliasing can be reduced by increasing the number of bits in the MISRs. However, in this section we introduce an alternative technique for reducing aliasing that does not require longer MISRs. This technique, which we call *Neighbor Consistency Checking (NCC)*, checks whether the trace generated by BMC corresponding to a particular candidate FF is consistent with the signatures captured in the MISRs at the interfaces of **other** blocks in the rest of the design. Any inconsistency can rule out the candidate FF as the one which captured the error, thereby improving diagnosis resolution.

The NCC strategy is as follows. Recall that for each FF candidate in the buggy block, the BMC tool also returns a full sequence of input and output values for *T* cycles (where *T* is the number of time frames analyzed). These values are applied as input or output constraints, as appropriate, to each of the neighboring blocks. The BMC tool is then used to check each neighboring block to see whether these newly-added constraints are consistent with the block's logic and the signatures on its interfaces. Note that, because the initial state of a neighboring block (initial states of FFs in the neighboring block) is not known, these checks cannot be done using simulation.

If a neighboring block is consistent, then the BMC tool returns a trace for inputs and outputs of that neighboring block. This sequence can then be recursively tested against additional neighbors, continuing across the chip. The limit would be reaching a clock domain boundary, where the interface is not shared directly with another block (as discussed in Sec. 2.1).

If the BMC tool reports an inconsistency while analyzing the **immediate** neighbors of a buggy block, the trace under consideration (and the corresponding candidate FFs) can be eliminated. If the analysis has progressed further, then we backtrack to the previously analyzed neighbor and check to see if a different trace can be found that is consistent with its signatures. We continue the analysis in this way until either a fully consistent (across the whole design) set of traces is found or no such set of traces can be found in which case the candidate trace is eliminated.

Let us return again to our running example. This time, we add a small neighboring block whose three inputs are driven by the three outputs (X, Y, Z) as shown in Fig. 10. The outputs of this block are captured using another 6-bit MISR (similar to Fig. 4). Recall from Sec. 2.3 (Fig. 8) that there are 4 FF candidates and 2 traces consistent with the signature from Output MISR 1 at cycle 6. If we now use those two traces as the input constraints on the neighboring block, we can test if they are consistent with Output MISR 2. As seen in Fig. 10, the output sequence for aliased Trace 2 (in Fig. 8) results in a different final signature in Output MISR 2. This means it can be ruled out as a potential candidate, reducing the candidate FFs to those from Trace 1 (in Fig. 8).

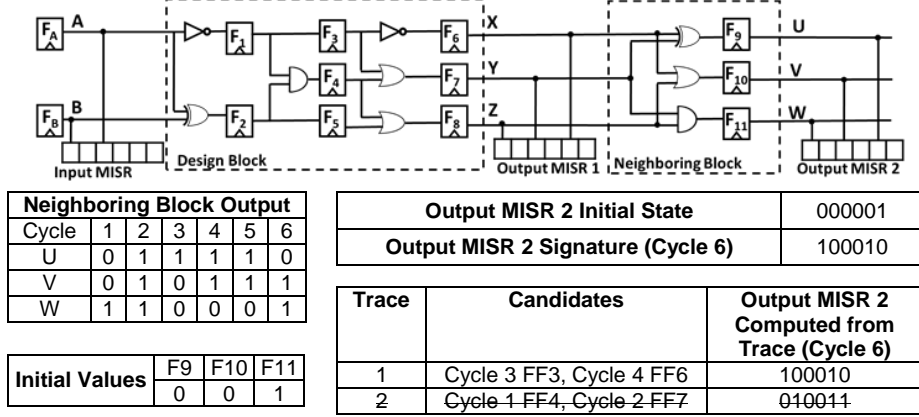

| Neighboring Block Output | | | | | | |
|---|---|---|---|---|---|---|
| Cycle | 1 | 2 | 3 | 4 | 5 | 6 |
| U | 0 | 1 | 1 | 1 | 1 | 0 |
| V | 0 | 1 | 0 | 1 | 1 | 1 |
| W | 1 | 1 | 0 | 0 | 0 | 1 |

| Initial Values | F9 | F10 | F11 |
|---|---|---|---|
| | 0 | 0 | 1 |

| Output MISR 2 Initial State | 000001 |
|---|---|
| Output MISR 2 Signature (Cycle 6) | 100010 |

| Trace | Candidates | Output MISR 2 Computed from Trace (Cycle 6) |
|---|---|---|
| 1 | Cycle 3 FF3, Cycle 4 FF6 | 100010 |
| ~~2~~ | ~~Cycle 1 FF4, Cycle 2 FF7~~ | ~~010011~~ |

**Fig. 10**: Example circuit (continuing from Fig. 8) demonstrating NCC.

## 3    Results

In this section, we present simulation results demonstrating the effectiveness and practicality of E-QED. For our simulations, we use the OpenSPARC T2 SoC [OpenSPARC], an open-source version of the UltraSPARC T2 SoC (a 500-million-transistor SoC with 8 processor cores with private L1 caches supporting 64 hardware threads, 8 banks of shared L2 cache memory using a directory-based cache coherence protocol, 4 on-chip memory controllers, and a crossbar-based interconnect). We used the following design parameters for the E-QED Signature Blocks (Sec. 2.1): target capture window $N = 1{,}024$ cycles, counter size $C = 11$ bits; MISR size $K = N * b$, where $b = 8$ for all MISRs except those inside the processor core, where $b = 4$. A total of 118 E-QED Signature Blocks were used, adding 185,057 FFs, with an area impact of 2.5%.

In addition to an overview of results (in Sec. 3.1) with the parameters listed above, we also present an analysis of various trade-offs (in Sec. 3.2) that influence total area impact (by varying the parameter *b* in various signature blocks and also the signature block locations) and the granularity of electrical bug localization (in terms of candidate FFs as well as the number of candidate traces). The results in this section are for the single-cycle FF error model of electrical bugs (Sec. 2.3).

### 3.1 Overview of Results

We randomly injected single-cycle FF errors in the OpenSPARC T2 design (with E-QED signature blocks inserted using our algorithm in Sec. 2.1) through RTL simulations (using Synopsys VCS). We ran 1,000,000 simulation runs, half with the SPLASH-2 Fast Fourier Transform (FFT) benchmark [Woo 95] and the other half with an in-house version of the Matrix Multiply (MMULT) benchmark. We ran 64-threaded versions of both benchmarks to ensure all threads on all processor cores of Open-SPARC were active. The single-cycle FF errors were injected into: the processor core (all instances), crossbar, L2 cache controller (all instances) and the memory control unit (all instances). For each simulation run, we ran both the original benchmark and the benchmark transformed using QED (EDDI-V and PLC transforms with *Inst_min* and *Inst_max* set to 100, see [Lin 14]). During each simulation run, the injected error was allowed to propagate until it either vanished (i.e., was masked before impacting any program outputs) or resulted in observable failure (e.g., crash, incorrect outputs, exception or error detection by QED versions of the benchmarks). As soon as a QED test detected an error, we performed electrical bug localization using E-QED (Fig. 2). For BMC, we used the Questa Formal tool (version 10.5) from Mentor Graphics on an Intel Xeon E5-2640 with 128GB of RAM.

Table 1 summarizes the results. The bug trace size is the number of signals for all input and output interfaces captured in the signature blocks for that design block, multiplied by the length of the trace generated by the BMC tool. Recall that this length is determined by the counter value ($T$ in Sec. 2.2) used to set up BMC analysis.

**Table** 1: Electrical bug localization results for 1,000,000 single-cycle FF error simulations in the OpenSPARC T2 SoC.

|  | Original tests (No QED) | E-QED (after running QED versions of the original tests) |
|---|---|---|
| Number of Detected Errors | 2,832 | 12,555 |
| Error Detection Latency (cycles) [Min, Avg, Max] | [2K, 976K, 8.8M] | [19, 168, 763] |
| Number of FF Candidates [Min, Avg, Max] | $9.3 * 10^5$ (~all flip-flops in the design) | [5, 18, 36] |
| Number of Candidate Bug Traces [Min, Avg, Max] | N/A | [1, 1, 1] |
| Number of bits in a bug trace [Min., Avg, Max] | N/A | [109k, 483k, 1.74M] |
| BMC Runtime (minutes) | N/A | [420, 526, 772] |
| Total E-QED signature block area impact | N/A | 2.5% |
| Clock speed impact | N/A | ~ 0% |
| **Localization Factor [Min, Avg, Max]** | 1x | [25,837x, 51,674x, 186,025x] |

In Table 1, the "Number of Detected Errors" is the number of injected errors detected by the corresponding validation test (deadlocks and hangs caused by either test were not included). Note that, the vast majority of injected errors vanished (which is consistent with published literature [Cho 15, Sanda 08]). QED tests detected many more errors than the original tests (~4x more than original tests due to fine-grained QED checks, which is also consistent with published literature [Hong 10, Lin 14]).

**Observation 1:** E-QED automatically localized every error detected by the EDDI-V and PLC checks in the QED tests, and produced a list of 18 candidate FFs on average (out of a total of close to a million FF candidates) – a localization factor of over four orders of magnitude. These results were achieved with BMC runtime of 7-13 hours. Very importantly, in every case, E-QED was able to report a unique trace with an average trace size of 483 kbits.

Table 2 shows the breakdown of results with respect to various design blocks of the OpenSPARC T2 SoC and various techniques presented in Sec. 2.2, 2.3 and 2.4.

**Table** 2: Electrical bug localization results (similar to Table 1) with a detailed breakdown.

|  |  | Design Block Localization (Sec. 2.2) | Single-Cycle FF Error Localization (Sec 2.3) | **NCC Localization (Sec. 2.4)** |
|---|---|---|---|---|
| Processor Core (Core) | Number of Traces [Min, Avg, Max] | N/A | [1, 3, 8] | **[1, 1, 1]** |
|  | Number of FF Candidates [Min, Avg, Max] | 44,084 | [11, 54, 91] | **[8, 16, 29]** |
|  | Avg. Localization Factor | 22x | 17,224x | **58,133x** |
|  | BMC Runtime (minutes) [Min, Avg, Max] | [62, 74, 77] | [303, 416, 478] | **[49, 85, 114]** |
| L2 Cache (L2C) | Number of Traces [Min, Avg, Max] | N/A | [2, 14, 19] | **[1, 1, 1]** |
|  | Number of FF Candidates [Min, Avg, Max] | 31,675 | [28, 77, 154] | **[7, 19, 33]** |
|  | Avg. Localization Factor | 29x | 12,080x | **48,954x** |
|  | BMC Runtime (minutes) [Min, Avg, Max] | [51, 58, 64] | [468,504,535] | **[42,187,242]** |
| Crossbar (CCX) | Number of Traces [Min, Avg, Max] | N/A | [2, 12, 17] | **[1, 1, 1]** |
|  | Number of FF Candidates [Min, Avg, Max] | 41,521 | [31, 74, 130] | **[5,23, 36]** |
|  | Avg. Localization Factor | 22x | 12,569x | **40,440x** |
|  | BMC Runtime (minutes) [Min, Avg, Max] | [53, 61, 72] | [379,421,450] | **[54,142,205]** |
| Memory Control Unit (MCU) | Number of Traces [Min, Avg, Max] | N/A | [3, 19, 24] | **[1, 1, 1]** |
|  | Number of FF Candidates [Min, Avg, Max] | 18,068 | [21, 67, 143] | **[5, 11, 18]** |
|  | Avg. Localization Factor | 51x | 13,882x | **84,557x** |
|  | BMC Runtime (minutes) [Min, Avg, Max] | [35, 37, 41] | [315,387,428] | **[78,163,251]** |
| Total E-QED signature block area impact | | **2.5%** | | |
| Clock speed impact | | **~0%** | | |

**Observation 2:** Signature-based electrical bug localization at the design block level (Sec. 2.2) achieves a localization factor of 22-51x. Single-Cycle FF Error Localization improves the localization factor by another 200-1,000x. Neighbor Consistency Checking (NCC) further improves localization factor by another 5-10x (for an average overall localization factor of 50,000x) and produces just a single candidate trace.

## 3.2 Design Trade-offs

For the results in Tables 1 and 2, the total area impact of E-QED signature blocks is 2.5% (obtained through synthesis of the OpenSPARC T2 SoC using the Synopsys Design Compiler with the Synopsys EDK 32 nm library for standard cells and memories). In Fig. 11, we analyze the area impact vs. the granularity of electrical bug localization. We vary the MISR parameter $b$ (recall from Sec. 2.1 that the MISR size is $b*M$, where $M$ is number of signals captured by the signature) from 2 to 14.[5] For the entire design, this varies the total number of FFs in the signature blocks from 48,436 to 338,930. We use the number of candidate traces here for comparison since the primary aim of NCC is to reduce the number of candidate traces found (ideally to 1). Reducing the number of candidate traces also reduces the number of FF candidates.

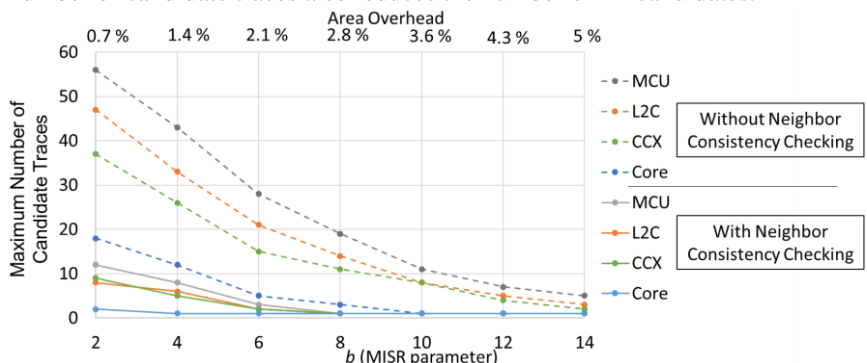

**Fig. 11**: The maximum number of unique candidate traces observed, before and after Neighbor Consistency Checking (NCC), for different area overheads, varied by changing the size of the MISRs in the signature blocks (MISR size is $b*M$, where $M$ is number of signals captured by the signature).

**Observation 3:** Neighbor Consistency Checking is able to reduce the number of candidate traces to just 1 when the area impact is 2.8% (this corresponds to a design where MISRs are constructed with $b = 8$ for the entire design). The area impact can be further reduced to 2.5% (while still ensuring only a single candidate trace is found) by using a hybrid set of parameters (the parameters used in Sec. 3.1), namely $b = 4$ within the processor cores and $b = 8$ everywhere else.

As discussed in Sec. 2.1, we made two enhancements to our approach to signature block insertion: signature blocks were added for signals around the L1 and L2 cache memory array banks (adding 16,640 FFs for all L2 banks and 3,840 FFs for all L1 banks), and signature blocks were also added at the instruction fetch units of the processors (adding 8,192 FFs across all 8 cores). As shown in Fig. 10, NCC analysis with these additional E-QED signature blocks significantly reduces the number of candidate traces for each value of $b$.

---

[5] Note that in our simulation experiments, above (in Sec. 3.1), we used different fixed values of $b$ for different parts of the design. In this section, we allow $b$ to vary but all MISRs in the design use the same value of $b$.

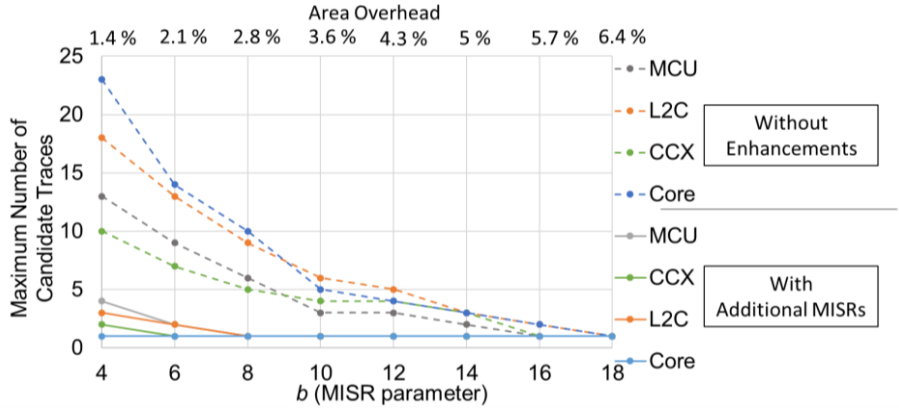

**Fig. 12**: The maximum number of unique candidate traces observed after performing NCC, with and without the addition of extra signature blocks around the L2 and L1 caches and processor instruction fetch unit, for different MISR parameters $b$ (From Sec. 2.1, MISR size is $b*M$, where $M$ is number of signals captured).

## 4 Related Work

Most existing publications on post-silicon bug localization (especially for electrical bugs) focus on ways to improve observability as well as techniques for localizing bugs inside the design (and combinations thereof). For example, many existing techniques capture system behavior (logic values of various signals during post-silicon validation runs) using trace buffers: small on-chip memories that record logic values of a selected set of internal signals [Abramovici 06, Anis 07, Ma 15, and many others]. However, with typical long error detection latencies (millions or billions of clock cycles) trace buffers can quickly become ineffective. E-QED leverages the QED technique [Lin 14] to create post-silicon validation tests with sufficiently short error detection latencies. (Other techniques that create tests with short error detection latencies can also be used with E-QED). Even when tests with short error detection latencies are available, inserting trace buffers for fine-grained electrical bug localization still imposes unacceptably high area overheads. For example, as discussed in Sec. 1, saving a full trace of all inputs and outputs of the crossbar block alone of the OpenSPARC T2 SoC for just 1,000 cycles would require over 34 Mbits of data. Recorders that store microarchitectural information in processor cores (e.g., [Park 09, 10]) require much less area. But they are applicable for processor cores only (while uncore components and accelerators occupy large portions of complex SoCs). Moreover, approaches in [Park 09, 10] require manual analysis. Techniques for compressing trace buffer contents generally provide limited benefits. Trace compression techniques such as state restoration [Ko 09] (or related approaches that use RTL simulation to restore signals from partial traces) cannot be used for electrical bugs (because electrical bugs are not present in the RTL description). [Vali 16] focuses presents a trace signal selection approach that directly tries to improve the ability to diagnose FF errors caused by electrical bugs. However, the reported improvements are modest.

Signature blocks used by E-QED overcome these limitations. Although MISRs have been extensively used for manufacturing testing (e.g., [Saxena 97 and numerous others]), E-QED does not require the generation of fault-free signatures by simulating post-silicon validation tests (unlike manufacturing testing). Instead, it uses formal techniques to reason about consistency between signatures captured by E-QED signature blocks at the inputs and outputs of a design block, as well as the logic inside the design block itself. While it may be possible to use hardware-implemented assertions (e.g., [Bayazit 05]) to enhance E-QED, finding the "right" set of (hardware implementation-friendly) assertions remains a major challenge for automatic assertion generation. Other observability techniques (e.g., [DeOrio 11, Li 10]) can result in false positives.

The problem of electrical bug localization is similar to the classical fault diagnosis problem in manufacturing testing [Abramovici 90]. The use of Scan Design for Testability makes the problem somewhat simpler for manufacturing testing vs. post-silicon validation. Many publications have used formal techniques for these purposes (e.g., [Larrabee 92, Mangassarian 07, and others]). Similarly, [Zhu 11, 14] use formal techniques, aided by backbones and a sliding window approach, for electrical bug localization. Techniques such as BackSpace and its derivatives [De Paula 08, 11, 12, Sengupta 12, Le 13] also use formal methods for bug localization purposes. However, as explained in Sec. 2, the biggest challenge is to create an automatic approach that scales for large SoCs without incurring large area impact. Some of the above techniques attempt to overcome the scalability challenge by "consistently" reproducing the buggy behavior over multiple runs; this can be very difficult for complex designs, as explained in [De Paula 11]. Our E-QED approach overcomes these challenges: E-QED signature blocks enable scalability for large SoC designs with small area impact; short error detection latencies of QED tests enable us to apply BMC in conjunction with the E-QED signature blocks; and, E-QED analysis techniques in Sec. 2 enable us to perform bug localization despite lossy compression by E-QED signature blocks (with minimal reliance on consistent reproduction of buggy behaviors). Note that, the E-QED Neighbor Consistency Checking technique in Sec. 2.4 is different from consistency checking techniques presented in [Jones 05, Park 09, 10, and others].

## 5 Conclusion

E-QED overcomes electrical bug localization challenges during post-silicon validation and debug. It automatically localizes electrical bugs and provides a comprehensive list of components that may contain the bugs (together with corresponding bug traces). It is an automatic approach which is highly effective and practical for large designs, as demonstrated on the OpenSPARC T2 SoC: automatic electrical bug localization in a few hours (9 hours on average) that can narrow an electrical bug to a handful of candidate flip-flops (18 flip-flops on average for a design with ~ 1 Million flip-flops) and identify a single candidate bug trace. The area impact of E-QED is ~2.5%. In contrast, it might take weeks (or even months) of mostly manual work (per bug) using traditional approaches. E-QED is made possible through a unique combination of Quick Error Detection techniques for bug detection, E-QED signature blocks that are automatically inserted during design, and new consistency checking techniques enabled by formal methods.

There are several future directions. E-QED can be extended to: 1. leverage already-existing Scan Design for Testability techniques to further enhance bug localization; 2. localize bugs in in analog and mixed-signal components of SoCs; 3. understand the interplay between scalability of BMC tools for bug localization vs. error detection latencies of QED tests vs. design of more sophisticated E-QED signature blocks; 4. diagnose defects that are detected using system-level testing during manufacture; 5. enable full system-level (consisting of many ICs chips) bug localization; and, 6. correct/fix bugs after manufacture.

## References

1. [Abramovici 90] M. Abramovici, M. A. Breuer, and A. D. Friedman, *Digital Systems Testing and Testable Design*, Computer Science Press, 1990.
2. [Abramovici 06] M. Abramovici, "A Reconfigurable Design-for-Debug Infrastructure for SoCs," in *Proceedings of IEEE/ACM Design Automation Conference*, pp. 7-12, 2006.
3. [Anis 07] E. Anis and N. Nicolici, "On Using Lossless Compression of Debug Data in Embedded Logic Analysis," in *Proceedings of 2007 IEEE International Test Conference (ITC)*, 2007.
4. [Bardell 87] P.H. Bardell, W.H. McAnney, J. Savir, *Built-in test for VLSI: Pseudorandom Techniques*, John Wiley and Sons, New York, 1987
5. [Bayazit 05] A. A. Bayazit and S. Malik, "Complementary use of runtime validation and model checking," in *Proceedings of ICCAD-2005. IEEE/ACM International Conference on Computer-Aided Design*, 2005, pp. 1052-1059.
6. [Cho 15] H. Cho, et al., "Understanding soft errors in uncore components," in *Proceedings of 2015 52nd ACM/EDAC/IEEE Design Automation Conference (DAC)*, pp. 1-6, 2015.
7. [Clarke 01] E. Clarke, A. Biere, R. Raimi, Y. Zhu, "Bounded Model Checking Using Satisfiability Solving," *Formal Methods in System Design*, Vol. 19, No. 1, pp. 7-34, 2001.
8. [DeOrio 11] A. DeOrio, D. S. Khudia and V. Bertacco, "Post-silicon bug diagnosis with inconsistent executions," in *Proceedings of 2011 IEEE/ACM International Conference on Computer-Aided Design (ICCAD)*, San Jose, CA, 2011, pp. 755-761
9. [De Paula 08] F. M. De Paula et al., "BackSpace: Formal Analysis for Post-Silicon Debug," in *Proceedings of Intl. Conference on Formal Methods in Computer-Aided Design*, pp. 1-10, 2008.
10. [De Paula 11] F. M. De Paula, et al., "TAB-BackSpace: Unlimited-Length Trace Buffers with Zero Additional On-Chip Overhead," in *Proceedings of IEEE/ACM Design Automation Conference*, 2011.
11. [De Paula 12] F. M. De Paula, A.J. Hu, and A. Nahir, "nuTAB-BackSpace: Rewriting to Normalize Non-Determinism in Post-Silicon Debug Traces," in *Proceedings of International Conference on Computer Aided Verification*, pp. 513-531, 2012.
12. [Dusanapudi 15] M. Dusanapudi et al., "Debugging post-silicon fails in the IBM POWER8 bring-up lab," *IBM Journal of Research and Development*, vol. 59, no. 1, pp. 12:1-10, Jan.-Feb. 2015.
13. [Foster 15] Foster, H. D., "Trends in Functional Verification: A 2014 Industry Study," *Proc. IEEE/ACM Design Automation Conference*, pp. 48-52, 2015.
14. [Friedler 14] O. Friedler et al., "Effective Post-Silicon Failure Localization Using Dynamic Program Slicing," *Proceedings of IEEE/ACM Design Automation Test in Europe*, pp. 1-6, 2014.
15. [Hong 10] T. Hong, et al., "QED: Quick Error Detection Tests for Effective Post-Silicon Validation," in *Proceedings of IEEE International. Test Conference,* pp. 1-10, 2010.
16. [Jones 05] R. B. Jones, C.-J. H. Seger, D. L. Dill, "Self-Consistency Checking," in *Proceedings of International Conference on Formal Methods in Computer-Aided Design*, pp. 159-171, 2005.
17. [Ko 09] H. F. Ko and N. Nicolici, "Algorithms for State Restoration and Trace-Signal Selection for Data Acquisition in Silicon Debug," *IEEE Transactions on Computer-Aided Design of Integrated Circuits and Systems*, vol. 28, no. 2, pp. 285-297, Feb. 2009.
18. [Larrabee 92] T. Larrabee, "Test pattern generation using Boolean satisfiability," *IEEE Transactions on Computer-Aided Design of Integrated Circuits and Systems*, vol. 11, no. 1, Jan 1992.

19. [Le 13] B. Le, D. Sengupta, A. Veneris and Z. Poulos, "Accelerating post silicon debug of deep electrical faults," in *Proceedings of 2013 IEEE 19th International On-Line Testing Symposium* (IOLTS), Chania, 2013, pp. 61-66.
20. [Li 10] W. Li, A. Forin, and S.A. Seshia. "Scalable Specification Mining for Verification and Diagnosis", in *Proceedings of Design Automation Conference (DAC)*, pp. 755–760, 2010.
21. [Lin 14] D. Lin, et al, "Effective Post-Silicon Validation of System-on-Chips Using Quick Error Detection," *IEEE Transactions on Computer Aided Design of Integrated Circuits Systems*, Vol. 33, No. 10, pp. 1573-1590, 2014.
22. [Lin 15] D. Lin, et al., "A Structured Approach to Post-Silicon Validation and Debug Using Symbolic Quick Error Detection," in *Proceedings of 2015 IEEE International Test Conference (ITC)*, Oct. 2015
23. [Lu 82] D. J. Lu, "Watchdog Processors and Structural Integrity Checking," in *IEEE Transactions on Computing.*, vol. 31, no. 7, pp. 681-685, Jul. 1982.
24. [Ma 15] S. Ma, et al., "Can't see the forest for the trees: State restoration's limitations in post-silicon trace signal selection", in *Proceedings of 2015 IEEE/ACM International Conference on Computer-Aided Design (ICCAD)*, pp. 1-8, 2015.
25. [Mangassarian 07] H. Mangassarian, et al., ``A Performance-Driven QBF-Based Iterative Logic Array Representation with Applications to Verification, Debug and Test," in *Proceedings of International Conference on Computer-Aided Design (ICCAD)*, 2007.
26. [McLaughlin 09] R. McLaughlin, S. Venkataraman and C. Lim, "Automated Debug of Speed Path Failures Using Functional Tests," in *Proceedings of 2009 IEEE VLSI Test Symposium*, pp. 91-96, 2009.
27. [Mishra 17] P. Mishra; R. Morad, A. Ziv, S. Ray, "Post-silicon Validation in the SoC Era: A Tutorial Introduction," in *IEEE Design & Test*, April 2017
28. [Nahir 14] A. Nahir, et al., "Post-Silicon Validation of the IBM POWER8 Processor," *Proceedings of IEEE/ACM Design Automation Conference*, pp. 1-6, 2014.
29. [Oh 02] Oh, N., et al.,"Control Flow Checking by Software Signatures," in *IEEE Transactions on Reliability*, Vol. 51, No. 1, pp. 111-122, Mar. 2002.
30. [OpenSPARC] *OpenSPARC: World's First Free 64-bit Microprocessor* [Online]. Available: http://www.opensparc.net
31. [Park 09] S.-B. Park, T. Hong, and S. Mitra, "Post-silicon Bug Localization In Processors Using Instruction Footprint Recording and Analysis (IFRA)," *IEEE. Transactions on Computer Aided Design Integrated Circuits System*, Vol. 28, No. 10, pp. 1545–1558, 2009.
32. [Park 10] S.-B. Park, *et al.,* "BLoG: Post-Silicon Bug Localization in Processors Using Bug Localization Graph", in *Proceedings of IEEE/ACM Design Automation Conference,* pp. 368-373, 2010.
33. [Reick 12] K. Reick, "Post-Silicon Debug – DAC Workshop on Post-Silicon Debug: Technologies, Methodologies, and Best-Practices,"in *Proceedings of IEEE/ACM Design Automation Conference,* 2012
34. [Sanda 08] P. N. Sanda et al., "Soft-error resilience of the IBM POWER6 processor," *IBM Journal of Research and Development*, vol. 52, no. 3, pp. 275–284, May 2008.
35. [Saxena 97] N. R. Saxena and E. J. McCluskey, "Parallel signature analysis design with bounds on aliasing," in *IEEE Transactions on Computers*, vol. 46, no. 4, pp. 425-438, Apr 1997.
36. [Sengupta 12] D. Sengupta, et al. "Lazy suspect-set computation: Fault diagnosis for deep electrical bugs.*"* in *Proceedings of the Great Lakes Symposium on VLSI*. ACM, 2012.
37. [Shen 83] J. P. Shen, M. A. Schuette, "On-line Self-Monitoring Using Signatured Instruction Streams," in *Proceedings of IEEE International Test Conference*, pp. 275–282, 1983.
38. [Shiu 01] P. H. Shiu, Tan Y., Mooney, V. J. "A novel parallel deadlock detection algorithm and architecture."in *Proceedings of the Ninth International Symposium on Hardware/Software Codesign, 2001*. IEEE, 2001.

39. [Singh 17] E. Singh, C. Barrett, S. Mitra, "E-QED: Electrical Bug Localization During Post-Silicon Validation Enabled by Quick Error Detection and Formal Methods," *International Conference on Computer-Aided Verification*, 2017
40. [Vali 16] A. Vali and N. Nicolici, "Bit-flip detection-driven selection of trace signals," in *Proceedings of 2016 21th IEEE European Test Symposium (ETS)*, Amsterdam, 2016, pp. 1-6
41. [Vermeulen 14] B. Vermeulen and K. Goossens. 2014. *Debugging Systems-on-Chip: Communication-Centric and Abstraction Based Techniques*. Springer.
42. [Woo 95] S. C. Woo et al., "The SPLASH-2 Programs: Characterization and Methodological Considerations*,"* in *Proceedings of International Symposium on Computer Architecture*, 1995
43. [Zhu 11] C.S. Zhu, G. Weissenbacher, and S. Malik, "Post-Silicon Fault Localisation Using Maximum Satisfiability and Backbones," in *Proceedings of IEEE/ACM Formal Methods Computer-Aided Design*, pp. 63-66, 2011.
44. [Zhu 14] C. S. Zhu, G. Weissenbacher and S. Malik, "Silicon fault diagnosis using sequence interpolation with backbones," in Proceedings of 2014 *IEEE/ACM International Conference on Computer-Aided Design (ICCAD),* San Jose, CA, 2014